\documentclass{article}
\usepackage{witmse,amsmath,epsfig,graphicx,epsfig,amsmath,amsbsy,amssymb,fullpage,citesort,verbatim}
\usepackage[usenames]{color}

\def \replevels{{\mathcal{R}}}

\title{INFORMATION COMPLEXITY AND ESTIMATION}
\name{Dror Baron$^1$}
\address{$ ^1$Department of Electrical and Computer Engineering,
  North Carolina State University, \\
  Raleigh, NC 27695, USA, barondror@ncsu.edu }
\begin{document}

\maketitle
\begin{abstract}
We consider an input $x$ generated by an unknown stationary ergodic source $X$ 
that enters a signal processing system $J$, resulting in $w=J(x)$.
We observe $w$ through a noisy channel, $y=z(w)$; 
our goal is to estimate $x$ from $y$, $J$, and knowledge of
$f_{Y|W}$.
This is  {\em universal} estimation, because $f_X$ is unknown.
We provide a formulation that describes a trade-off between information complexity and noise.
Initial theoretical, algorithmic, and experimental evidence is presented in support of our approach.
\end{abstract}
%
\section{INTRODUCTION}
\label{sec:intro}

{\em Universal} algorithms
\cite{LZ77,Rissanen1984,Rissanen1984,SingerFeder1998,BaronWeissman2011submit,Jalali2008}
achieve the best possible performance asymptotically -- without knowing the input statistics.
These algorithms have had tremendous impact in lossless compression,
which is crucial for data backups and transmissions.
In sharp contrast, universal algorithms have made much less impact in other areas.

\begin{figure}[t]
\vspace*{-22mm}
\begin{center}
\includegraphics[width=85mm]{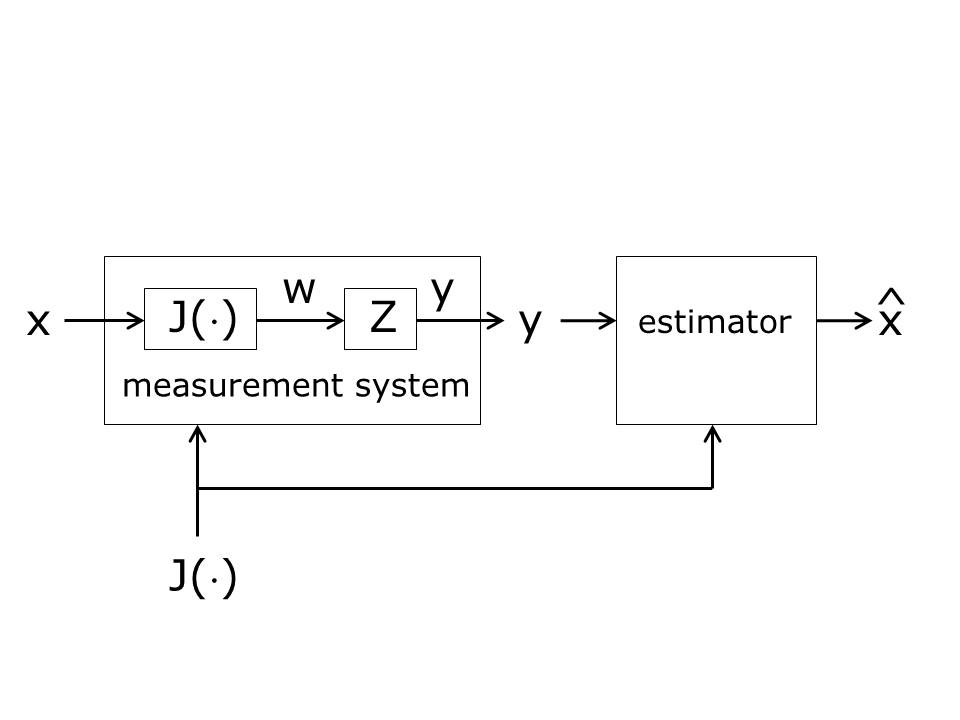}
\end{center}
\vspace*{-15mm}
\caption{
{\small\sl 
{\bf Measurement and estimation system}:\
An {\em input} $x\in\mathbb{R}^N$ 
generated by an unknown stationary ergodic source $X$
is processed by a known (potentially nonlinear) operator $J$
to produce $w=J(x)\in\mathbb{R}^L$. A probabilistic {\em noise} operator $z$ 
that implies a known probability density $f_{Y|W}(y|w=J(x))$ is applied to $w$, the
measurements are $y = z(J(x))$.
Our goal is to estimate $x$ using $y \in \mathbb{R}^M$ and $J$,
resulting in $\widehat{x}\in\mathbb{R}^N$. 
Although our emphasis is on real-valued $w,x,y$, 
discrete-valued signals and operators are allowed.
}
\label{fig:system}
}
\end{figure}

{\em Estimation} algorithms attempt to recover an input from
noisy measurements (Figure~\ref{fig:system}).
Numerous estimation problems have received great attention including
the additive noise scalar channel, 
$y=x+z$~\cite{Cover06};
linear matrix multiplication with additive noise, $y=Jx+z$
with applications including  compressed sensing
\cite{CandesRUP,DonohoKolmogorovCS2006,HN05},
finance,  medical and seismic imaging;
universal lossy compression~\cite{Jalali2008,BaronWeissman2011submit,Yang1999},
where the goal is to find compressible $x$ that is sufficiently close to $y$;  
nonlinear regression, where $J(x)$ is nonlinear; and
distributed signal processing.

In these estimation problems, the common goal is to 
estimate the input $x$ from knowledge of the
noisy measurements $y$ and measurement system $J$.
To do so, we must exploit all statistical structure in $x$.
A particularly challenging type of statistical structure is the appearance of 
spatial or temporal dependencies in data.
In images, such dependencies can be captured by dictionary learning
or employing energy compacting transforms.
In other problems, the statistical dependencies might be more subtle.
Following the lead of universal lossless compression, we assume that the input $x$ was generated by 
an unknown stationary ergodic source $X$.
It is well known that stationary ergodic 
models capture the statistics of text files well, and hence the 
success of universal lossless compressors. 
Stationary ergodic models have also been incorporated in speech denoising
and enhancement, and appear prominently 
in hidden Markov models.

One approach to {\em universal estimation} relies on Kolmogorov 
complexity~\cite{Kolmogorov1965}.
For a prospective $\widehat{x}$, the Kolmogorov complexity
$K(\widehat{x})$ is the length of the shortest computer program that can
compute $\widehat{x}$. 
Donoho~\cite{DonohoKolmogorov} proposed a Kolmogorov-based
estimator for the white scalar channel, $y=x+z$.
Despite related extensions to compressed 
sensing~\cite{DonohoKolmogorovCS2006,HN05},
{\em what is missing in the literature is a universal approach in arbitrary
measurement systems
that would support noise and unknown stationary ergodic 
input distributions. }

We propose to perform universal estimation in (potentially nonlinear) signal processing
systems from noisy measurements.
The algorithmic component of our work features  a harmonious marriage of scalar quantization, 
universal lossless compression, and Markov chain Monte Carlo. 
We evaluate the estimated input $\widehat{x}$ over a quantized grid and optimize for 
the trade-off 
between information complexity (lossless coding length) of $\widehat{x}$ and 
how well $\widehat{x}$ explains the measurements $y$.
We report promising preliminary theoretical and numerical results.

\vspace*{-0.25mm}
\section{Information complexity formulation}
\vspace*{-0.25mm}
\label{sec:theory}

We focus on the setting where the lengths 
$M$ of the output $y$ and
$N$ of the input $x$
both grow to infinity, $M,N\rightarrow\infty$.
We further assume that their ratio is finite and positive, 
$\lim_{N\rightarrow\infty} \frac{M}{N} = \delta>0$.
Similar settings have been discussed in the literature,  
e.g.,~\cite{GuoVerdu2005}.
Since $x$ was generated by an unknown source, we must search for an estimation 
mechanism that is agnostic to the specific distribution $f_X$. 

{\bf Kolmogorov complexity:}\ 
For $x \in \mathbb{R}^N$, the Kolmogo-rov complexity~\cite{Kolmogorov1965}
of $x$, denoted by $K(x)$, is the length of the shortest computer program that can
compute $x$. To be more precise, $K(x)$ is the length of the shortest input to a 
Turing machine~\cite{Turing1950} that generates $x$ and then halts.
We limit our discussion to Turing machines whose ``input tapes'' consist of bits.
Consider the shortest program $\mathcal{P}(x)$ that generates $x$. From the 
perspective of a source encoder~\cite{Cover06},  we say that $\mathcal{P}(x)$ is a code for $x$. 

Having linked Turing machines~\cite{Turing1950} and data 
compression~\cite{Cover06}, let us {\em temporarily} limit the discussion 
to {\em discrete valued} $x$ 
generated by a stationary ergodic source $X$. Each such $x$ is generated with 
probability $p_X(x)$, and it is easily shown that the per-symbol Kolmogorov coding length
$K(x)$ converges to the entropy rate $H$ almost surely, 
$\lim_{N\to\infty} \frac{1}{N}K(x)=H$~\cite{Cover06}.
Noting that universal lossless compressors~\cite{LZ77,Rissanen1984} 
achieve $H$ asymptotically for discrete 
valued stationary ergodic sources~\cite{Cover06},
we see that these algorithms achieve 
the per-symbol Kolmogorov complexity almost surely.

{\bf Kolmogorov sampler:}\
For additive white Gaussian noise,
$y=x+z$,
Donoho \cite{DonohoKolmogorov} proposed the 
{\em Kolmogorov sampler},
\[
\widehat{x}_{KS}
={\arg\min}_{\hat{x}} \{ K(\hat{x}) - \log(f_{Z}(z=y-\hat{x}))\}.
\]
For stationary ergodic $X$, $\widehat{x}_{KS}$ is sampled from
the posterior $f_{X|Y}(y|x)$, where the mean square error,
$E[(\widehat{x}_{KS}-x)^2]$,
is twice larger than the 
Bayesian minimum mean square error (MMSE)~\cite{DonohoKolmogorov}.

In a later paper, Donoho et al. discussed a Kolmogorov estimator
for compressed sensing $y=Jx$~\cite{DonohoKolmogorovCS2006};
their estimator ignores noise, and is of limited practical interest.
For the noisy version of this problem, $y=Jx+z$,
Haupt and Nowak~\cite{HN05} described a complexity
measure that, when optimized, produces the LASSO algorithm~\cite{Tibshirani1996}. 
To the best of our knowledge, Haupt and Nowak did not pursue complexity based
regularization beyond iid signals and additive white Gaussian noise (AWGN).

{\bf Quantization and estimation:}\ 
The overwhelming majority of real numbers have infinite Kolmogorov complexity.
Nonetheless, some scalars $x\in \mathbb{R}^N$ can be represented by a finite length 
$\mathcal{P}(x)$.
In practice, it is impossible to compute $K(x)$ even for discrete alphabets.
At the same time, we have seen that universal lossless source 
codes~\cite{LZ77,Rissanen1984}
achieve per-symbol Kolmogorov coding length almost surely~\cite{Cover06}.
To represent continuous valued $\widehat{x}$,
we apply a scalar quantizer,
$Q:\widehat{x}\in\mathbb{R}^N \longrightarrow x' \in Q^N$, 
and then compress $x'=Q(\widehat{x})$
with a universal lossless compressor $U$ with coding length
$U(x')$, where quantization levels $Q\subset \mathbb{R}$ 
consist of a {\em finite} subset of $\mathbb{R}$, 
and performing an optimization over $\widehat{x} \in Q^N$
reduces the complexity of the estimation problem from infinite to combinatorial. 
Note that we generate $x'$ by 
independently quantizing each entry of $x$ with $Q$.
This {\em encoder} first describes the quantizer $Q$ and 
then compresses $Q(x)$. The coding length, which we desire to minimize,
is denoted by $U(Q(x))$ or $U(x)$.

It would seem that we must search for a good scalar quantizer $Q$ 
(Section~\ref{sec:algos}), but  
{\em data-independent} reproduction levels are of theoretical interest,
\begin{equation*}
\replevels \triangleq \left\{
-\frac{\gamma^2}{\gamma},-\frac{\gamma^2-1}{\gamma},\ldots,
\frac{\gamma^2}{\gamma}
\right\}, \quad \gamma=\lceil\log(N)\rceil.
\end{equation*}
As $N$ increases, $\replevels$ will quantize a 
broader range of values of $x$ to a greater resolution. 
An encoder based on $\replevels$ need not describe the structure of the data-independent quantizer,
because $N$ is known. That is, $U(\replevels(x))$ only accounts for the length of
the universal code $U$. 

{\bf Universal MAP estimation:}\
We perform  maximum {\em a posteriori} (MAP)
estimation over possible sequences  $\widehat{x}\in\replevels^N$,
where the prior $p_X(x)=2^{-U(\widehat{x})}$ 
utilizes the coding length $U(\widehat{x})$ of some universal
lossless compressor~\cite{LZ77,Rissanen1984},
\begin{equation}
\label{eq:x_MAP_Kolmogorov}
\widehat{x}_{MAP} = \arg\min_{\hat{x}\in\replevels^N} \{ U(\hat{x}) - \log(f_{Y|W}(y|w=J(\hat{x}))) \}, 
\end{equation}
where we note that $\replevels(\hat{x})=\hat{x}$ for $\hat{x}\in\replevels^N$.
Our MAP estimator is applicable 
to {\em any} signal processing system $J$ and supports {\em any} probabilistic noise 
operators, it is closely related to universal 
prediction~\cite{SingerFeder1998,Rissanen1984}.

{\bf Estimation performance:}\
We have promising preliminary theoretical results using the data-independent quantizer $\replevels$.
In universal lossy source coding of analog (continuous valued) 
sources~\cite{BaronWeissman2011submit}, we have shown with Weissman
that $\widehat{x}_{MAP}$ (\ref{eq:x_MAP_Kolmogorov})
achieves the rate distortion function for finite variance stationary ergodic sources
in an appropriate asymptotic sense.
That is, $U(\replevels(\widehat{x}))$ offers a sufficiently good approximation to $K(\widehat{x})$ 
in universal lossy compression,
where we chose $U(\widehat{x})$ to be empirical entropy
of blocks of $q=O(\log(N))$ symbols in $\widehat{x}$.
In universal compressed sensing~\cite{BaronDuarteAllerton2011}, 
we have shown with Duarte
that under minor technical conditions on $f_X$, performing MAP estimation over the discrete 
alphabet $\replevels$ 
converges to the MAP estimate over the continuous distribution $f_X$ asymptotically, 
where we used
i.i.d. zero-mean Gaussian noise $z\in\mathbb{R}^M$ with known variance. 
It remains to be seen whether $\replevels$ or other data-independent
quantizers are useful for {\em arbitrary} nonlinear measurement systems.

\begin{figure}[t]
\vspace*{-3mm}
\begin{center}
\hspace*{-2mm}
\includegraphics[width=90mm]{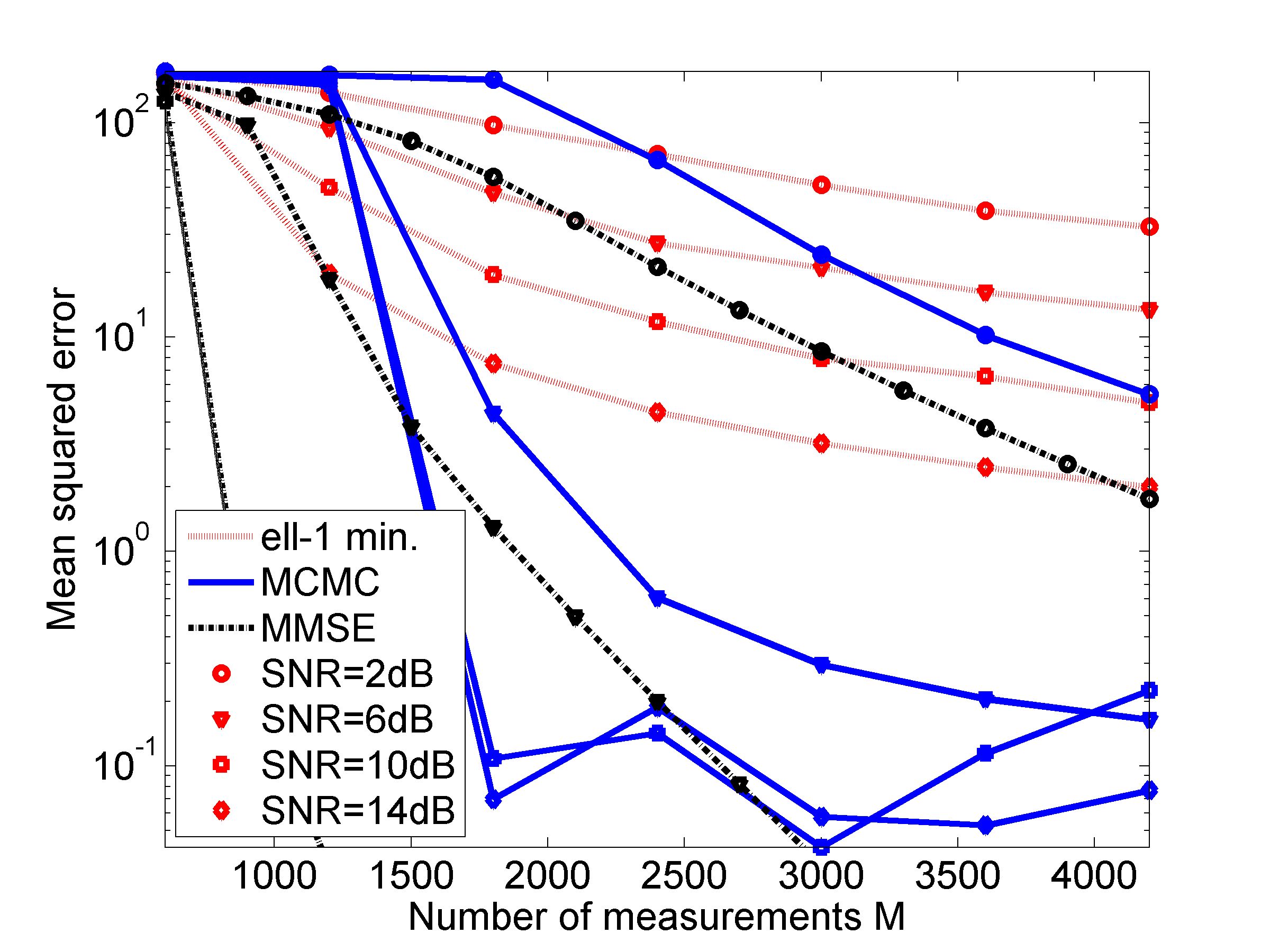}
\end{center}
\vspace*{-5mm}
\caption{
{\small\sl 
Universal Markov chain Monte Carlo (MCMC)~\cite{BaronDuarteAllerton2011} 
and $\ell_1$-norm minimization~\cite{CandesRUP}
recovery results for a source with 
i.i.d.\ Bernoulli entries with nonzero probability of 3\% as a function of the number of 
Gaussian random measurements $M$ for different signal to noise ratio (SNR) values. }  
\label{fig:universal_CS_Bernoulli}
}
\end{figure}

In terms of the mean square error,
we would expect $\widehat{x}_{MAP}$ to perform well
in Donoho's scalar channel setting, $y=x+z$. 
With Duarte~\cite{BaronDuarteAllerton2011}, we have promising results for the compressed 
sensing (linear matrix multiplication) channel,
$y=Jx+z$, where we approximated $\widehat{x}_{MAP}$  (\ref{eq:x_MAP_Kolmogorov})
by a Markov chain Monte Carlo (MCMC)~\cite{Geman1984} algorithm (Section~\ref{sec:algos}).
Figure~\ref{fig:universal_CS_Bernoulli} illustrates recovery results from 
Gaussian measurement matrices for a source with 
i.i.d.\ Bernoulli entries with nonzero probability of 3\%.
Our MCMC algorithm outperforms $\ell_1$-norm minimization, which is a well-known
compressed sensing reconstruction (estimation) algorithm \cite{CandesRUP},
except when the number of measurements $M$ is low. 
Comparing MCMC to the minimum mean square error (MMSE) 
achievable in the Bayesian regime with known statistics~\cite{GuoVerdu2005},
the square error achieved by MCMC is {\em three} times larger. 
One is left to wonder whether 
the mean square error performance of our algorithm might also be double the MMSE, 
particularly in the limit of infinite computation (Section~\ref{sec:algos}).

{\bf Taking Kolmogorov beyond MAP:}\
The Kolmogorov sampler $\widehat{x}_{KS}$
samples from the posterior~\cite{DonohoKolmogorov};
it throws away all the statistical information it has on signals $\widehat{x}$ that 
differ from $\widehat{x}_{KS}$.
Seeing that the mean square error obtained by $\widehat{x}_{KS}$ is double the MMSE,
there is great potential to reduce estimation error over our Kolmogorov-based 
MAP estimator  $\widehat{x}_{MAP}$ (\ref{eq:x_MAP_Kolmogorov}).
We therefore propose Kolmogorov-based
conditional expectation,
\begin{eqnarray*}
\widehat{x}_{MSE} &=& 
 E[x|J,y] \\
&=&
\frac{
\sum_{\hat{x}\in\replevels^N} \hat{x} \cdot 2^{-U(\hat{x})} f_{Y|W}(y|w=J(\hat{x})) }
{ \sum_{\hat{x}\in\replevels^N} 2^{-U(\hat{x})} f_{Y|W}(y|w=J(\hat{x}))},
\end{eqnarray*}
where we employ
the universal prior, $p_X(\hat{x})=2^{-U(\replevels(\hat{x}))}$.
It is well known that conditional expectation achieves the MMSE
of the Bayesian regime, and this estimator should perform well.
Interestingly, when the signal to noise ratio (SNR) is low,
the Bayesian MMSE is sizable, and achieving double the MMSE 
is unimpressive. In these low SNR settings, $\widehat{x}_{MSE}$ should 
estimate {\em much} better than $\widehat{x}_{MAP}$.

In some signal processing systems,
one wants to minimize some other (not necessarily quadratic) distortion metric
$D(x,\widehat{x})$. The universal prior is readily invoked
by defining the Kolmogorov conditional probability,
\[
p_{X|Y}(x|y)  = \frac{ p_{Y|X} p_X}{ p_Y} \propto p_{Y|X} p_X,
\]
and taking the minimizing expression gives the Kolmogorov-based estimator
for $D(\cdot)$,
\begin{equation*}
\widehat{x}_D = \arg\min_w \left\{ \sum_{\widehat{x}\in\replevels^N} D(\widehat{x},w)f_{Y|W}(y|w=J(\hat{x})) 2^{-U(\hat{x})} \right\}.
\end{equation*}
For scalar channels and iid noise, Sivaramakrishnan and Weissman~\cite{Sivaramakrishnan2008}
described a universal denoising algorithm that estimates $x$ by $\widehat{x}_{SW}$, 
its expected error $E[D(x,\widehat{x}_{SW})]$ converges to the Bayesian risk 
asymptotically in an appropriate stochastic setting.
For scalar channels and iid noise,
our expected estimation error $E[D(x,\widehat{x}_D)]$ should also be
asymptotically optimal. 
The performance in arbitrary signal processing systems $J$ is an open question.

\vspace*{-0.25mm}
\section{Algorithms}
\label{sec:algos}
\vspace*{-0.25mm}
\label{sec:algos}

In principle,  $\widehat{x}_{MAP}$ can be computed by evaluating
the Kolmogorov-based posteriors of
$|\replevels|^N$ possible sequences $\replevels(x)$. This is better than 
continuous estimation, but still computationally intractable.
Instead, we perform this optimization using Markov chain Monte Carlo (MCMC)~\cite{Geman1984,Jalali2008},
where  $U(\widehat{x})=H_q(\widehat{x})$ is the empirical entropy
of blocks of $q=O(\log(N))$ symbols of $\widehat{x}$.

{\bf Markov chain Monte Carlo:}\
We use MCMC~\cite{Geman1984} to approximate
$\widehat{x}_{MAP}$, which is the globally optimal MAP minimizer. 
To keep things simple, assume that $\widehat{x}\in\replevels^N$
is a candidate estimate. Define the Boltzmann PDF,
\begin{equation}
\label{eq:def_Boltzmann}
f_s(\widehat{x}) \triangleq \frac{1}{\zeta_s} \exp(-s [  H_q(\widehat{x}) - \log(f_{Y|W}(y|w=J(\widehat{x})))  ]),
\end{equation}
where $H_q(x)$ is the empirical entropy of blocks of $q$ symbols in 
$x$~\cite{Rissanen1984,Jalali2008,BaronWeissman2011submit,BaronDuarteAllerton2011},
$q=O(\log( N))$ to ensure convergence of the empirical entropy to the entropy rate~\cite{Cover06},
$s>0$ is inversely related to temperature in an analogous statistical physics heat-bath setting~\cite{Geman1984},
and $\zeta_s$ is a normalization constant. 
To sample from the Boltzmann PDF (\ref{eq:def_Boltzmann}), we use a 
{\em Gibbs sampler}: in each iteration, a single element $\widehat{x}_n$ is generated 
by resampling from the PDF, while the rest of $\widehat{x}$ remains unchanged.  
The key idea is to reduce temperatures slowly enough for the
randomness of Gibbs sampling to eventually drive MCMC out of any local 
minimum toward the globally optimal $\widehat{x}_{MAP}$. 

{\bf Adaptive quantizer:}\
Jalali and Weissman~\cite{Jalali2008}
have used MCMC to approach the fundamental rate distortion (RD)
limits~\cite{Cover06} in lossy compression of binary inputs.
For continuous valued (analog) 
sources~\cite{BaronWeissman2011submit},
using the data-independent quantizer $\replevels$ 
in MCMC asymptotically achieves the RD function universally for 
stationary ergodic continuous amplitude sources. However, $\replevels$ grows with 
the input length, slowing down the convergence to the RD function, 
and is thus an impediment in practice.

To address this issue,  we next propose an MCMC-based algorithm 
that uses an
{\em adaptive quantizer} $Q$. The ground-breaking work by Rose on
the discrete nature of the Shannon codeboook
for iid sources when the Shannon
lower bound is not tight~\cite{Rose94} suggests that, for most sources
of practical interest, restriction of the quantizer $Q$ to
a smaller number of levels does not stand in the way of attaining the
fundamental compression limits.
When employed on such sources, our latter algorithm zeroes in on the finite
quantizer, and thus enjoys rates of convergence commensurate 
with the small-quantizer setting.

\begin{figure}[t]
\vspace*{-9mm}
\begin{center}
\hspace*{-15mm}
\includegraphics[width=200mm]{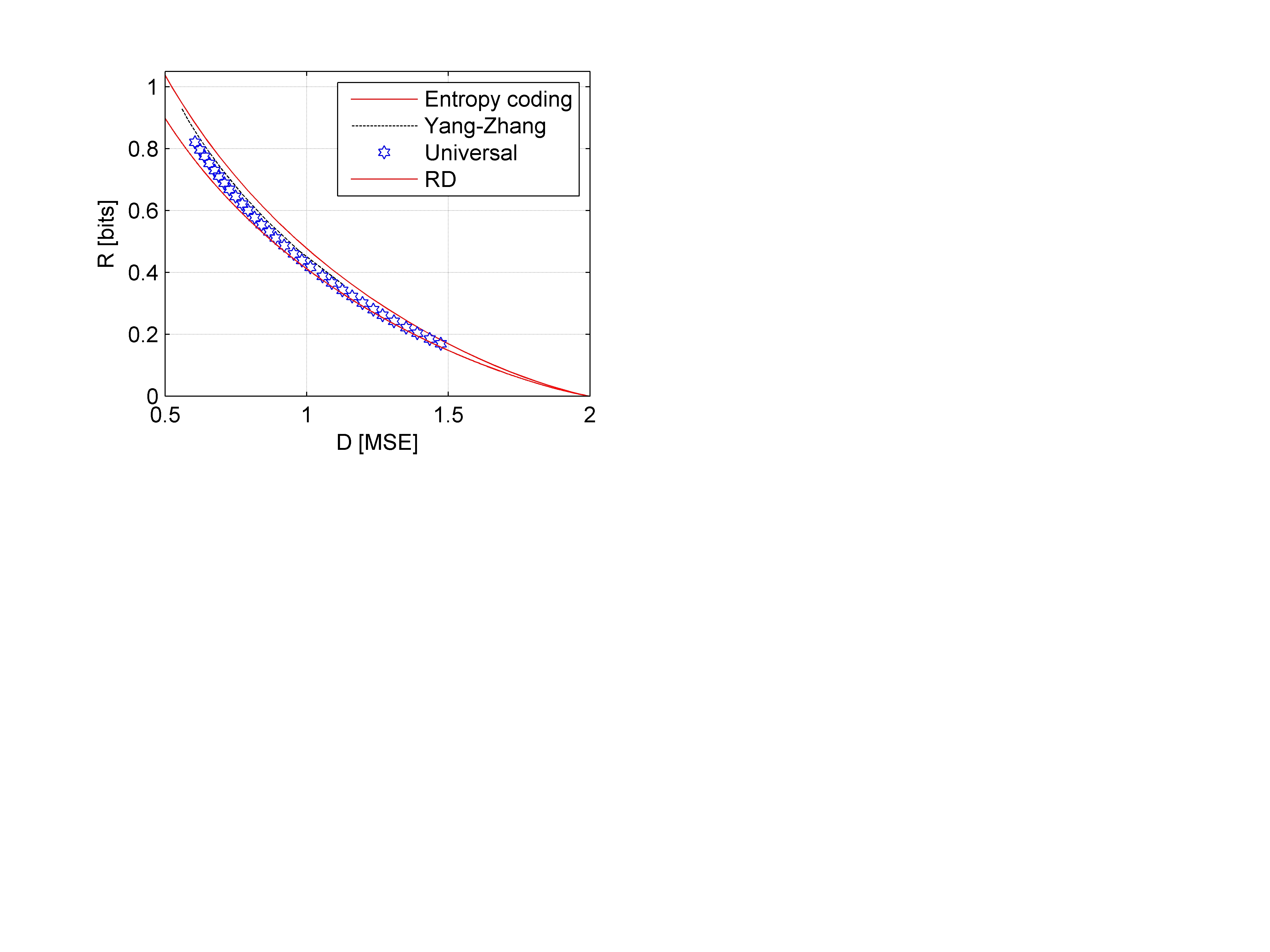}
\end{center}
\vspace*{-83mm}
\caption{
{\small\sl {\bf Universal lossy compression}:\
Rate $R$ vs. distortion $D$ of entropy coding~\cite{GershoGray1993},
results by Yang and Zhang~\cite{Yang1999},
average rate and distortion of our universal lossy compression 
algorithm~\cite{BaronWeissman2011submit},
and the RD function~\cite{Cover06} for length-15000 iid Laplace inputs, 
$f_X(x)=\frac{1}{2}e^{-|x|}$. }
\label{fig:MCMC_Laplace}
}
\end{figure}

{\bf Numerical results:}\
In universal lossy compression of analog sources~\cite{BaronWeissman2011submit},
we have developed an algorithm that optimizes the quantizer for square error,
and have promising preliminary results.
Figure~\ref{fig:MCMC_Laplace} compares results for an iid Laplace input, $f_X(x)=\frac{1}{2}e^{-|x|}$,
achieved by entropy coding~\cite{GershoGray1993},
a deterministic approach by Yang and Zhang \cite{Yang1999},
and our universal MCMC algorithm~\cite{BaronWeissman2011submit}.

In our universal compressed sensing work with Duarte \cite{BaronDuarteAllerton2011}, 
we focused on development of
a fast routine for optimizing the quantizer; this routine greatly accelerates 
the algorithm.
We have seen in Figure~\ref{fig:universal_CS_Bernoulli}
for a source with i.i.d.\ Bernoulli entries with nonzero probability of 3\%
that MCMC outperforms $\ell_1$-norm minimization, 
except when the number of measurements $M$ is low.
We have additional results, but omit these for brevity;
MCMC generally estimates the input signal $x$ well, but much
work remains to be done. 

\section{ACKNOWLEDGMENTS}

I wish to thank Marco Duarte and Tsachy Weissman for permission to discuss our joint
work~\cite{BaronWeissman2011submit,BaronDuarteAllerton2011}.
Thanks also to Neri Merhav, Deanna Needell, and
Phil Schniter for enlightening discussions.

\bibliographystyle{IEEEbib}
\small
\bibliography{cites}

\end{document}